\documentclass{article}
\usepackage{newtxmath}       
\usepackage{textcomp}
\usepackage{graphicx,amscd,amsmath,amssymb,verbatim}
\usepackage{textcomp}
\usepackage[OT1,T1]{fontenc}
\usepackage{url}

\newtheorem{theorem}{Theorem}

\begin{document}
%
\title{Projective representations of the inhomogeneous symplectic group: Quantum symmetry origins of the Heisenberg commutation relations}
%
%
\author{Stephen G. Low}
%
%
%
\maketitle              

\begin{abstract}
Quantum symmetries that leave invariant physical transition probabilities are described by projective representations of Lie groups. The mathematical theory of projected representations for topologically connected Lie groups is reviewed and applied to the inhomogeneous symplectic group $\mathcal{I}\mathcal{S}p(2n)$. The projective representations are given in terms of the ordinary unitary representations of the central extension of the group with respect to its first homotopy group direct product with its second cohomology group. The second cohomology group of $\mathcal{I}\mathcal{S}p(2n)$ is the one dimensional abelian group and the central extension turns the $2n$ dimension abelian normal subgroup of translations of  
$\mathcal{I}\mathcal{S}p(2n)$ into the Weyl-Heisenberg group. This is the quantum symmetry origin of the Heisenberg commutation relations that are the Hermitian representation of the Lie algebra of the Weyl-Heisenberg group.
\end{abstract}

\section{Projective representations}
A fundamental quantum property is that the physical measurable transition
probabilities are given by the square of the modulus of the states.
This results in a quantum phase that is responsible for quantum
phenomena such as interference. Physical measurable transition probability
from state $\alpha $ to $\beta $ is given by,
\begin{equation}
P( \alpha \rightarrow \beta ) =|\left( \Psi _{\beta },\Psi _{\alpha
}\right) |^{2}.%
\label{transprob}
\end{equation}

The physical states $\Psi $ are rays that are equivalence classes
of states in a Hilbert space $\text{\boldmath $\mathrm{H}$}$ defined
up to a phase, $\Psi =[|\psi \rangle ]=\{e^{i \theta }\overset{
}{|\psi \rangle }\}$, $|\psi \rangle \in \text{\boldmath $\mathrm{H}$}$,
$\theta \in \mathbb{R}$. Suppose that you have a representation
$\upsilon$ of a Lie group $\mathcal{G}$ on $\text{\boldmath $\mathrm{H}$}$
such that $\upsilon ( g) :\text{\boldmath $\mathrm{H}$}\rightarrow
\text{\boldmath $\mathrm{H}$}:\tilde{|\psi \rangle }\mapsto \upsilon
( g) \overset{ }{|\psi \rangle }$, $g\in \mathcal{G}$. This induces
transformations on the rays $\tilde{\Psi }=\upsilon ( g)  \Psi $. Then
$\mathcal{G}$ is a symmetry group and $\upsilon $ is called a projective
representation if it preserves the transition probabilities,
\begin{equation}
P( \alpha \rightarrow \beta ) =|\left( {\tilde{\Psi }}_{\beta },{\tilde{\Psi
}}_{\alpha }\right) |^{2}=|\left( \upsilon ( g)  \Psi _{\beta },\upsilon
( g)  \Psi _{\alpha }\right) |^{2}=|\left(  \Psi _{\beta },\Psi
_{\alpha }\right) |^{2}.%
\label{transprobprojrep}
\end{equation}

These projective representations that describe quantum symmetries
also have a phase degree of freedom that are responsible for many
of their uniquely quantum properties. Several basic theorems enable
us to compute projective representations for symmetry groups of
physical interest. First, Wigner \cite{wigner}, showed
that any projective representation acting on a separable Hilbert
space is equivalent to a representation that is either linear and
unitary or anti-linear and anti-unitary (see Appendix B of Chapter
2 in \cite{Weinberg1}). As a Lie group that is topologically
path connected to the identity does not admit anti-linear, anti-unitary
representations, projective representations of connected Lie groups
are always equivalent to a representation that are linear and unitary. The
anti-linear, anti-unitary representations only are required for
groups such as the extended Lorentz group $\mathcal{I}\mathcal{O}(
1,n) $ that has disconnected components. 

The inhomogeneous symplectic group that is of interest in this paper
is topologically path connected and so we restrict to the case of
connected Lie symmetry groups that considerably simplifies the analysis. We
then have the fundamental theorem that is due to the work of Bargmann
\cite{bargmann} and Mackey \cite{mackey2},\cite{mackey}:
\begin{theorem} The projective representations
of a topologically path connected Lie group {\upshape $ \mathcal{G}$}
acting on a separable Hilbert space $\text{\boldmath $\mathrm{H}$}$
are equivalent to the ordinary unitary representations of its unique
maximal central extension $\check{\mathcal{G}}$.\label{Theorem:projrep}
\end{theorem}

The essentials of the proof are as follows. The central extension
$\check{\mathcal{G}}$ of a group $\mathcal{G}$ is defined by the
short exact sequence, $e\rightarrow \mathcal{Z}\overset{i}{\rightarrow
}\check{\mathcal{G}}\overset{\rho }{\rightarrow }\mathcal{G}\rightarrow
e$. $e$ is the trivial group and as the sequence is exact, $i,\rho
$ are homomorphisms such that $\operatorname{image}( i) \simeq \ker
( \rho ) $. As the extension is central, $\mathcal{Z}$ is an abelian
group that is injected into the center of $\check{\mathcal{G}}$, $i(
\mathcal{Z}) \subset \check{\mathcal{G}}$. 

The unitary representations of an abelian group are its characters
that are given by phases. Therefore, for the unitary representations
of the group $\mathcal{G}$, its center is realized as phases. These
phases are normed to unity by the square of the modulus in the computation
of the transition probabilities (\ref{transprob}). From this,
it is straightforward to show that if ordinary unitary representations
of a symmetry group $\mathcal{G}$ satisfy (\ref{transprobprojrep}),
then the unitary representations of any central extension of $\check{\mathcal{G}}$
will also satisfy (\ref{transprobprojrep}). Conversely, if the
ordinary unitary representations of a centrally extended group $\check{\mathcal{G}}$
satisfies (\ref{transprobprojrep}), then the ordinary unitary
representations of any group $\mathcal{G}$ that it is a central
extension of will also satisfy (\ref{transprobprojrep}). Note
that the complete set of unitary irreducible representations of
a central extension $\check{\mathcal{G}}$ includes the representations
of $\mathcal{G}$ that is homomorphic to it as degenerate representations.

The central extension with respect to $\mathcal{Z}$ is maximal if
any other admissible central extension with respect to $\tilde{\mathcal{Z}}$
is such that $\tilde{\mathcal{Z}}\subset \mathcal{Z}$. This unique
maximal central extension exists for topologically connected Lie
groups \cite{Azcarraga}: 
\begin{theorem} The unique maximal
central extension $\check{\mathcal{G}}$ of a topologically path
connected Lie group $\mathrm{\mathcal{G}}$ exists and is given by
the short exact sequence\label{Theorem:projrepisunitaryCE}
\begin{equation}
 e\rightarrow \pi _{1}( \mathcal{G}) \otimes H^{2}( \mathcal{G},\mathbb{R})
\overset{i}{\rightarrow }\check{\mathcal{G}}\overset{\mathrm{\rho
}}{\rightarrow }\mathrm{\mathcal{G}}\rightarrow e 
\end{equation}where $\pi _{1}( \mathcal{G}) $ is the fundamental
homotopy group of {\upshape $ \mathcal{G}$} and $H^{2}( \mathcal{G},\mathbb{R})
$ is the second cohomology group of $\mathrm{\mathcal{G}}$. Consequently,
$\check{\mathcal{G}}$ is connected and simply connected.   
\end{theorem}

Note that $\mathbb{A}\simeq \pi _{1}( \mathcal{G}) $ is a discrete
(i.e finite or countably infinite) abelian group and $\mathcal{A}\simeq
H^{2}( \mathcal{G},\mathbb{R}) $ is a continuous group that is isomorphic
to $\mathcal{A}( m) $ for some $m$ where $\mathcal{A}( m) $ is
the continuous abelian group given by the reals under addition, $\mathcal{A}(
m) \equiv (\mathbb{R}^{m},+)$. The central extension with respect
to just the homotopy group $\pi _{1}( \mathcal{G}) $ is referred
to as the topological central extension. For a connected Lie group,
this defines the unique universal covering lie group $\overline{\mathcal{G}}$,
$e\rightarrow \pi _{1}( \mathcal{G}) \rightarrow \overline{\mathrm{\mathcal{G}}}\rightarrow
\mathrm{\mathcal{G}}\rightarrow e$. 

The central extension with respect to just the second cohomology
group is referred to as the algebraic central extension. It is a
local property that is completely defined by the Lie algebra. Suppose
that the Lie algebra of $\mathcal{G}$ has a basis $\{X_{a}\}$, $a,b=1,..
\dim  (\mathcal{G})$, that satisfy the commutation relations
$[X_{a},X_{b}]=c_{a,b}^{c}X_{c}$. Then consider all possible
central generators $\{I_{\alpha }\}$,
\begin{equation}
\left[ X_{a},X_{b}\right] =c_{a,b}^{c}X_{c}+d_{a,b}^{\alpha }I_{\alpha
}, \ \ \left[ X_{a},I_{\alpha }\right] =0,\  \ \left[ I_{\alpha },I_{\beta
}\right] =0.
\end{equation}

These must satisfy the Jacobi identities which constrains admissible
central generators and trivial extensions that are just translations
of the generators $X_{a}\rightarrow X_{a}+I_{a}$ are discarded.
The remaining $m$ nontrivial generators $\{I_{\alpha }\}$, $\alpha
=1,..,m$ are the generators of the abelian group that is the second
cohomology group, $\mathcal{A}( m) \simeq H^{2}( \mathcal{G},\mathbb{R})
$. The unique universal covering group with this Lie algebra
is the central extension $\check{\mathcal{G}}$.

\subsection{Central extension examples}

This method may be straightforwardly applied to groups of interest. 

First, for the abelian group $\mathcal{A}( n) $, $\pi _{1}( \mathcal{A}(
n) ) =e$ and $H^{2}( \mathcal{A}( n) ,\mathbb{R}) \simeq \mathcal{A}(
\frac{n(n-1)}{2}) $. This follows directly by noting that the
generators $A_{i}$ of the Lie algebra of $\mathcal{A}( n) $, $[A_{i},A_{j}]=0$,
$i,j =1,...,n$, admit $\frac{n(n-1)}{2}$ nontrivial central generators
$I_{i,j}$, $i<j$ that satisfy $[A_{i},A_{j}]=I_{i,j}$. 

Next, we consider the connected component of the inhomogeneous Lorentz
group. We denote $\mathcal{L}( 1,n) $ to be the orthochronus
Lorentz group that is the connected subgroup of $\mathcal{O}( 1,n)
$. The inhomogeneous Lorentz group $\mathcal{I}\mathcal{L}( 1,n)
\simeq \mathcal{L}( 1,n) \ltimes\mathcal{A}( n+1) $ has a trivial
second cohomology group, $H^{2}( \mathcal{I}\mathcal{L}( 1,n)
,\mathbb{R}) =e$. This calculation is given in \cite{Weinberg1},\cite{Azcarraga}.
The Jacobi identities involving the Lorentz generators constrain
the admissible nontrivial central generators of the abelian normal
subgroup to zero. The fundamental homotopy group of $\mathcal{L}(
1,n) $ is $\mathbb{Z}_{2}$ and so $\pi _{1}( \mathcal{I}\mathcal{L}(
1,n) ) =\mathbb{Z}_{2}$. The unique maximal central extension is
the cover, 
\begin{equation}
 e\rightarrow \mathbb{Z}_{2}\rightarrow \mathcal{S}pin( 1,n) \ltimes
\mathcal{A}( n+1) \rightarrow \mathcal{I}\mathcal{L}( 1,n) \rightarrow
e,
\end{equation}

\noindent where $\mathcal{S}pin( 1,n) \simeq \overline{\mathcal{L}}(
1,n) $ and, in particular, $\mathcal{S}pin( 1,3) \simeq \mathcal{S}\mathcal{L}(
2,\mathbb{C}) $. 

Finally, the inhomogeneous symplectic group $\mathcal{I}\mathcal{S}p(
2n) \simeq \mathcal{S}p( 2n) \ltimes\mathcal{A}( 2n) $ is a
connected Lie group. It has a nontrivial second cohomology group
that is a one dimensional abelian group, $H^{2}( \mathcal{I}\mathcal{S}p(
2n) ,\mathbb{R}) \simeq \mathcal{A}( 1) $; the central extension
of the Lie algebra admits the single nontrivial central generator
$I$ satisfying the commutation relations $[A_{i},A_{j}]=\delta _{i,j}I$
where $A_{i}$ are the generators of the abelian $\mathcal{A}( 2n)
$ subgroup. This calculation is given in \cite{Low}. The
unique simply connected universal covering group with this Lie algebra
is the Weyl-Heisenberg group $\mathcal{H}( n) $. \ The fundamental
homotopy group of $\mathcal{S}p( 2n) $ is $\mathbb{Z}$ and so $\pi
_{1}( \mathcal{I}\mathcal{S}p( 2n) ) =\mathbb{Z}$. The unique maximal
central extension is, 
\begin{equation}
 e\rightarrow \mathbb{Z}\otimes \mathcal{A}( 1) \rightarrow \overline{\mathcal{H}\mathcal{S}p}(
2n) \rightarrow \mathcal{I}\mathcal{S}p( 2n) \rightarrow e,%
\label{CEofisp}
\end{equation}

\noindent where $\mathcal{H}\mathcal{S}p( 2n) \simeq \mathcal{S}p(
2n) \ltimes\mathcal{H}( n) $ and where $\mathcal{H}( n)
\simeq \mathcal{A}( n) \ltimes\mathcal{A}( n+1) $. The full
central extension that is the universal cover is of the form 
\begin{equation}
\overline{\mathcal{H}\mathcal{S}p}( 2n) \simeq \overline{\mathcal{S}p}(
2n) \ltimes\mathcal{H}( n) .
\end{equation}

\subsection{Unitary representations of central extensions}

Levi's decomposition theorem \cite{barut} states that any
Lie algebra may be uniquely decomposed into a semidirect sum of
a semisimple algebra and a radical that is the maximum solvable
ideal and this is unique up to equivalence. A direct corollary is
that the unique simply connected Lie group with this Lie algebra
is always equivalent to a semidirect product of a simply connected
semi-simple Lie subgroup $\mathcal{K}$ and a simply connected solvable
normal Lie subgroup $\mathcal{N}$. As the central extension of
a connected Lie group is simply connected, it always is of the form
of this semidirect product, 
\begin{equation}
\check{\mathcal{G}}\mathrm{\simeq }\mathcal{K}\ltimes\mathcal{N}
.%
\label{CEsemidirect}
\end{equation}

Whitehead's lemma \cite{barut} establishes that the second
cohomology group of any semisimple group is trivial, $H^{2}( \mathcal{K},\mathbb{R})
\simeq e$, and so $H^{2}( \mathcal{G},\mathbb{R}) \subset
\mathcal{N}$.

\subsubsection{Unitary representations of semidirect product groups:}

The semidirect product structure of the central extension enables
us to make some general remarks about their unitary irreducible
representations before specializing to the inhomogeneous symplectic
group of primary interest in this paper. These unitary representations
have been studied by Mackey \cite{mackey2},\cite{mackey}
for a general class of locally compact topological groups. Our restriction
here to topologically connected smooth Lie groups radically simplifies
the treatment. 

We begin by reviewing the very well known special case of direct
products of the form $\mathcal{G}\simeq \mathcal{K}\otimes \mathcal{N}$
for which the elements satisfy $g=h k=k h$ where $g\in \mathcal{G},k\in
\mathcal{K}, h\in \mathcal{N}$. In this case, both subgroups
are normal and the inner automorphisms are trivial; $\tilde{h}=
k h k^{-1}=h$ and $\tilde{k}= h k h^{-1}=k$. \ The complete set
of unitary irreducible representations $\upsilon $ of $\mathcal{G}$
are 
\begin{equation}
\upsilon ( g) :{\text{\boldmath $\mathrm{H}$}}^{\upsilon }\rightarrow
{\text{\boldmath $\mathrm{H}$}}^{\upsilon },\ \   \upsilon =\sigma
\otimes \xi ,\ \    {\text{\boldmath $\mathrm{H}$}}^{\upsilon
}={\text{\boldmath $\mathrm{H}$}}^{\sigma }\otimes {\text{\boldmath
$\mathrm{H}$}}^{\xi } , \text{}
\end{equation}

\noindent where $\text{$\xi$}$ are unitary irreducible representations
of $\mathcal{N}$ on ${\text{\boldmath $\mathrm{H}$}}^{\xi }$, $\xi
( h) :{\text{\boldmath $\mathrm{H}$}}^{\xi }\rightarrow {\text{\boldmath
$\mathrm{H}$}}^{\xi }$ and $\text{$\sigma$}$ are unitary irreducible
representations of $\mathcal{K}$ on ${\text{\boldmath $\mathrm{H}$}}^{\sigma
}$, $\sigma ( k) :{\text{\boldmath $\mathrm{H}$}}^{\sigma }\rightarrow
{\text{\boldmath $\mathrm{H}$}}^{\sigma }$. (We label the Hilbert
space with the representation as the representation and group determine
the Hilbert space.) Representations of automorphisms are trivial
$\xi ( \tilde{h}) =\upsilon (k)\xi ( h) \upsilon ( k^{-1}) \equiv
\xi ( h) $ and likewise for $\sigma $. 

For a semidirect product $\mathcal{G}\simeq \mathcal{K}\ltimes\mathcal{N}$
that is not a direct product, $\mathcal{N}$ is normal but $\mathcal{K}$
is not normal and the inner automorphisms $\tilde{h}=k^{-1}h
k$ are generally not trivial. Elements can always be decomposed
uniquely as $g= h k=k \tilde{h}$, $\tilde{h}=k^{-1}h k$.

The general form of the unitary irreducible representations are:
 
\begin{equation}
\upsilon ( g) :{\text{\boldmath $\mathrm{H}$}}^{\upsilon }\rightarrow
{\text{\boldmath $\mathrm{H}$}}^{\upsilon },\ \   \upsilon =\sigma
\otimes \rho ,\ \    {\text{\boldmath $\mathrm{H}$}}^{\upsilon
}={\text{\boldmath $\mathrm{H}$}}^{\sigma }\otimes {\text{\boldmath
$\mathrm{H}$}}^{\xi }, 
\end{equation}

\noindent where $\rho $ is a nontrivial representation of some $\mathcal{G}^{0}=\mathcal{K}^{0}\otimes
_{s}\mathcal{N}\subset \mathcal{G}$, with $\mathcal{K}^{0}\subset
\mathcal{K}$ such that
\begin{equation}
\rho ( g) :{\text{\boldmath $\mathrm{H}$}}^{\xi }\rightarrow {\text{\boldmath
$\mathrm{H}$}}^{\xi },\ \   g\in \mathcal{G}^{0},  \ \rho
|_{\mathcal{N}}=\xi  .
\end{equation}

\noindent Representations of the automorphisms are now generally
nontrivial,
\begin{equation}
\xi ( \tilde{h}) =\upsilon ( k) \xi ( h) {\upsilon ( k) }^{-1}=\rho
( k) \xi ( h) {\rho ( k) }^{-1}.%
\label{repaut}
\end{equation}

As the group is simply connected, its group elements are globally
given by the exponentials of the generators of the Lie algebra. Suppose $\{X_{a}\}$ are
a basis of the Lie algebra of $\mathcal{N}$ with and $\{Y_{\alpha
}\}$ are a basis of the Lie algebra of $\mathcal{K}$. Then,
$\rho ^{\prime }( X_{a}) =\xi ^{\prime }( X_{a}) $ span the Hilbert
space ${\text{\boldmath $\mathrm{H}$}}^{\xi }$. Note that a subset
of these generators $X_{i}=I_{i}$ may be central generators. Therefore,
by Shurr's lemma, their representation is a multiple of the identity,
$\xi ^{\prime }( X_{i}) =\xi ^{\prime }( I_{i}) =\lambda _{i}\text{\boldmath
$1$}$, $\lambda _{i}\in \mathbb{R}$. As the $\rho $ also act
on the Hilbert space ${\text{\boldmath $\mathrm{H}$}}^{\xi }$, its
representations $\rho ^{\prime }$ of the Lie algebra of $\mathcal{K}$
must be defined in terms of the enveloping algebra of the Lie algebra
of $\mathcal{N}$. That is, $\rho ^{\prime }( Y_{\alpha }) =p_{\alpha
}( \xi ^{\prime }( X_{a}) ) =\rho ^{\prime }( p_{\alpha }( X_{a})
) $ for some polynomial $p_{\alpha }$. For the faithful representation,
this means that $Y_{\alpha }=p_{\alpha }( X_{a}) $. This must
satisfy the Lie algebra of $\mathcal{K}$ up to factors of central
generators.  

\subsubsection{Abelian normal subgroup:}

The case where the normal subgroup $\mathcal{N}\simeq \mathcal{A}(
n) $ is an abelian group has been extensively studied. The enveloping
algebra of the Lie algebra of an abelian group is abelian and therefore
there the Lie algebra of a nonabelian group $\mathcal{K}$ cannot
be nontrivially realized in its enveloping algebra. Consequently,
the extension $\rho $ is trivial, $\rho ( k) \equiv 1$ for $k\in
\mathcal{K}$ and $\mathcal{K}^{0}=e$. Then, the automorphism
condition (\ref{repaut}) reduces to
\begin{equation}
\xi ( \tilde{h}) =\xi ( k h k^{-1}) =\rho ( k) \xi ( h) {\rho (
k) }^{-1}=\xi ( h)  \forall  h\in \mathcal{N}.
\end{equation}

\noindent This is the fixed point equation in the unitary dual $\text{\boldmath
$U$}( \mathcal{A}( n) ) \simeq \mathbb{R}^{*n}$ that defines the
little groups $\mathcal{K}\mbox{}^{\circ}$,
\begin{equation}
\xi ( k h k^{-1}) =\xi ( h) \ \   \forall  h\in \mathcal{N},\  k\in
\mathcal{K}\mbox{}^{\circ}.
\end{equation}

The unitary irreducible representations $\upsilon \mbox{}^{\circ}=\sigma
\mbox{}^{\circ}\otimes \xi $ may then be defined on the stabilizers
$\mathcal{G}\mbox{}^{\circ}\simeq \mathcal{K}\mbox{}^{\circ}\otimes
_{s}\mathcal{N}$. The Mackey Theorems for semidirect products with
abelian normal subgroups then state that the representations induced
on the full group by these representations of the stabilizer subgroups
defines the full set of unitary irreducible representations of $\mathcal{G}\simeq
\mathcal{K}\ltimes\mathcal{A}( n) $. 

\subsubsection{Perfect-Mackey group:}

Suppose $\mathcal{N}$ is nonabelian and the representation $\rho
$ exists globally on all of $\mathcal{G}$, 
\begin{equation}
\xi ( \tilde{h}) =\xi ( k h k^{1}) =\rho ( k) \xi ( h) {\rho ( k)
}^{-1}\ \   \forall  h\in \mathcal{N},\  k\in \mathcal{K}.%
\label{perfectmackeyaut}
\end{equation}

Then, the representations $\upsilon =\sigma \otimes \rho $ acting
on ${\text{\boldmath $\mathrm{H}$}}^{\upsilon }={\text{\boldmath
$\mathrm{H}$}}^{\sigma }\otimes {\text{\boldmath $\mathrm{H}$}}^{\xi
}$ are a complete set of faithful unitary irreducible representations
of $\mathcal{G}$. We call groups with this property {\itshape
perfect-Mackey} groups due to the simple form of the faithful representations.

\section{Projective representations of the inhomogeneous symplectic
group}

The inhomogeneous symplectic group is a smooth Lie group that is
also a matrix group and therefore we can apply Theorem \ref{Theorem:projrep}
to compute its projective representations in terms of the ordinary
unitary representations of its central extension. The central
extension of $\mathcal{I}\mathcal{S}p( 2n) $ is the group $\overline{\mathcal{H}\mathcal{S}p}(
2n) $ given in (\ref{CEofisp}). Note that this has the required
semidirect product structure $\mathcal{K}\ltimes\mathcal{N}$ given
in (\ref{CEsemidirect}) where $\mathcal{K}=\overline{\mathcal{S}p}(
2n) $ is the simply connected semi-simple group and $\mathcal{N}=\mathcal{H}(
n) $ is the simply connected solvable normal subgroup. 

\subsection{$\overline{\mathcal{H}\mathcal{S}p}( 2n) $ is a perfect-Mackey
group}

We establish that $\overline{\mathcal{H}\mathcal{S}p}( 2n) $ is
a perfect-Mackey group as described above by explicitly constructing
the $\rho ^{\prime }$ representation for all of $\overline{\mathcal{H}\mathcal{S}p}(
2n) $ \cite{Low}, \cite{Low13}. \ Let $Z_{\alpha
}$ be the generators of algebra of $\mathcal{H}( n) $ and $W_{\alpha
,\beta }$ the generators of the algebra of $\overline{\mathcal{S}p}(
2n) $, $\alpha ,\beta ,..=1,...,2n$. The Lie algebra of $\mathcal{H}(n)$
has nonzero commutators, 
\begin{equation}
\left[ Z_{\alpha },Z_{\beta }\right] =\zeta _{\alpha ,\beta }I,\ \ \ 
\mbox{{\textlbrackdbl}}\zeta _{\alpha ,\beta }\mbox{{\textrbrackdbl}}=\left(
\begin{array}{ll}
 0 & 1 \\
 -1 & 0
\end{array}\right) .
\end{equation}

The Hermitian representation of the generators are denoted ${\hat{Z}}_{\alpha
}=\rho ^{\prime }( Z_{\alpha }) =\xi ^{\prime }( Z_{\alpha }) $
and $\hat{I}=$$\rho ^{\prime }( I) =\xi ^{\prime }( I) $ that act
on the Hilbert space ${\text{\boldmath $\mathrm{H}$}}^{\xi }=L^{2}(
\mathbb{R}^{n},\mathbb{C}) $. As $I$ is a central generator,
by Shurr's lemma its representation is $\hat{I}=\lambda  \text{\boldmath
$1$}$ , $\lambda \in \mathbb{R}\backslash \{0\}$ where {\bfseries
1} is the identity operator on ${\text{\boldmath $\mathrm{H}$}}^{\xi
}$. Then the representation is the familiar Heisenberg commutation
relations 
\begin{equation}
\left[ {\hat{Z}}_{\alpha },{\hat{Z}}_{\beta }\right] =i \lambda
\zeta _{\alpha ,\beta }. %
\label{Heisenbergcomm}
\end{equation}

Define generators $W_{\alpha ,\beta }$ of $\overline{\mathcal{S}p}(
2n) $ in terms of elements in the enveloping algebra of the Weyl-Heisenberg
group, $W_{\alpha ,\beta }=\zeta _{\alpha ,\gamma }Z_{\gamma
}Z_{\beta }+\zeta _{\beta ,\gamma }Z_{\gamma }Z_{\alpha }$. The
commutation relations satisfy the Lie algebra of the symplectic
group up to a central factor $I$,
\begin{equation}
\begin{array}{l}
 \left[ W_{\alpha ,\beta },Z_{\kappa }\right] =I \left( \zeta _{\alpha
,\kappa }Z_{\beta }+\zeta _{\beta ,\kappa }Z_{\alpha }\right) ,
\\
 \left[ W_{\alpha ,\beta },W_{\kappa ,\epsilon }\right] =I( \zeta
_{\alpha ,\epsilon }W_{\beta ,\kappa }+\zeta _{\alpha ,\epsilon
}W_{\beta ,\kappa }+\zeta _{\alpha ,\epsilon }W_{\beta ,\kappa }+\zeta
_{\alpha ,\epsilon }W_{\beta ,\kappa }) .
\end{array}
\end{equation}

The $\rho ^{\prime }$ Hermitian representations of generators ${\hat{W}}_{\alpha
,\beta }=\zeta _{\alpha ,\gamma }{\hat{Z}}_{\gamma }{\hat{Z}}_{\beta
}+\zeta _{\beta ,\gamma }{\hat{Z}}_{\gamma }{\hat{Z}}_{\alpha }$ act
on the Hilbert space ${\text{\boldmath $\mathrm{H}$}}^{\xi }$,
\begin{equation}
\begin{array}{l}
  \left[ {\hat{W}}_{\alpha ,\beta },{\hat{Z}}_{\kappa }\right] =i
\lambda ( \zeta _{\alpha ,\kappa }{\hat{Z}}_{\beta }+\zeta _{\beta
,\kappa }{\hat{Z}}_{\alpha }) ,  \\
 \left[ {\hat{W}}_{\alpha ,\beta },{\hat{W}}_{\kappa ,\epsilon }\right]
=i \lambda ( \zeta _{\alpha ,\epsilon }{\hat{W}}_{\beta ,\kappa
}+\zeta _{\alpha ,\epsilon }{\hat{W}}_{\beta ,\kappa }+\zeta _{\alpha
,\epsilon }{\hat{W}}_{\beta ,\kappa }+\zeta _{\alpha ,\epsilon }{\hat{W}}_{\beta
,\kappa }) .
\end{array}%
\label{repofhspalg}
\end{equation}

\noindent Rescaling the generators ${\hat{Z}}_{\beta }\rightarrow
\frac{1}{\sqrt{\lambda }}{\hat{Z}}_{\beta }$ removes the $\lambda
$ and (\ref{Heisenbergcomm}) and (\ref{repofhspalg}) define
the required Hermitian representation $\rho ^{\prime }$ of the Lie
algebra of $\overline{\mathcal{H}\mathcal{S}p}( 2n) $ acting
on the Hilbert space ${\text{\boldmath $\mathrm{H}$}}^{\xi }$ to
establish that it is a perfect-Mackey group. \ 

\subsection{Unitary representations of $\overline{\mathcal{H}\mathcal{S}p}(
2n) $}

By Theorem \ref{Theorem:projrep}, the projective representations
of $\mathcal{I}\mathcal{S}p( 2n) $ are equivalent to the ordinary
unitary representation $\overline{\mathcal{H}\mathcal{S}p}( 2n)
$. As $\overline{\mathcal{H}\mathcal{S}p}( 2n) $ is a perfect-Mackey
group, its faithful unitary representations are $\upsilon =\sigma
\otimes \rho $. That is, for $g\in \mathcal{G}$, $h\in \mathcal{H}(
n) $, $k\in \overline{\mathcal{S}p}( 2n) $, $g=h k$, 
\begin{equation}
\upsilon ( g) =\upsilon ( h k) =\sigma ( k) \otimes \rho ( h k)
=\sigma ( k) \otimes \xi ( h) \rho ( k) .%
\label{unitrephsp}
\end{equation}

\noindent $\sigma $ are ordinary unitary irreducible representations
of $\overline{\mathcal{S}p}( 2n) $,
\begin{equation}
\sigma ( k) :{\text{\boldmath $\mathrm{H}$}}^{\sigma }\rightarrow
{\text{\boldmath $\mathrm{H}$}}^{\sigma },\  k\in \overline{\mathcal{S}p}(
2n) .%
\label{unitrepsp}
\end{equation}

\noindent $\xi $ are ordinary unitary irreducible representations
of $\mathcal{H}( n) $,
\begin{equation}
\xi ( h) :{\text{\boldmath $\mathrm{H}$}}^{\xi }\rightarrow {\text{\boldmath
$\mathrm{H}$}}^{\xi },\  h\in \mathcal{H}( n) .
\end{equation}

\noindent $\rho $ are the unitary representations of $\overline{\mathcal{S}p}(
2n) $ that satisfy (\ref{perfectmackeyaut}), 
\begin{equation}
\rho ( k) :{\text{\boldmath $\mathrm{H}$}}^{\xi }\rightarrow {\text{\boldmath
$\mathrm{H}$}}^{\xi },\  k\in \overline{\mathcal{S}p}( 2n) .%
\label{rhospbar}
\end{equation}

Note that the defining equations for the Weil metaplectic representation
are
\begin{gather}
\tilde{\rho }( k) :{\text{\boldmath $\mathrm{H}$}}^{\xi }\rightarrow
{\text{\boldmath $\mathrm{H}$}}^{\xi },\  k\in \mathcal{M}p( 2n),
\end{gather}
\begin{equation}
\xi ( \tilde{h}) =\xi ( k h k^{1}) =\tilde{\rho }( k) \xi ( h) {\tilde{\rho
}( k) }^{-1}\ \   \forall  h\in \mathcal{H}( n).
\end{equation}

\noindent where the metaplectic group $\mathcal{M}p( 2n) $ is the
double cover of the symplectic group satisfying the short exact
sequence 
\begin{equation}
e\rightarrow \mathbb{Z}_{2}\rightarrow \mathcal{M}p( 2n) \rightarrow
\mathcal{S}p( 2n) \rightarrow e.
\end{equation}

\noindent $\overline{\mathcal{S}p}( 2n) $ is the cover of $\mathcal{M}p(
2n) $ with the short exact sequence, $e\rightarrow \mathbb{Z}/\mathbb{Z}_{2}\rightarrow
\overline{\mathcal{S}p}( 2n) \overset{\pi }{\rightarrow }\mathcal{M}p(
2n) \rightarrow e$. Then $\rho =\tilde{\rho }\circ \pi $ is the
$\rho $ representation (\ref{rhospbar}) of $\overline{\mathcal{S}p}(
2n) $ that satisfies (\ref{perfectmackeyaut}). These representations
may be computed explicitly as described in Chapter 4 of \cite{folland}
and \cite{Low}.

In principle we can compute the ordinary unitary irreducible representations
$\sigma $ of the cover of the symplectic group (\ref{unitrepsp}). These
may be combined in (\ref{unitrephsp}) to give the faithful unitary
irreducible representations of the group $\overline{\mathcal{H}\mathcal{S}p}(
2n) $. The degenerate representations that are the unitary representations
of the groups that are homomorphic to $\overline{\mathcal{H}\mathcal{S}p}(
2n) $ must also be considered for a complete set of representations.
These include $\mathcal{I}\overline{\mathcal{S}p}( 2n) $ and $\overline{\mathcal{S}p}(
2n) $. The unitary representations of the inhomogeneous group
can be computed usuing the Mackey theory for abelian normal subgroups.  \ 

\section{Summary}

Our discussion started with physically measurable quantum transition
probabities that are given by the square of the modulus of states.
Quantum symmetries leaving the transition probabiities invariant
are described by projective representations of Lie groups. The projected
representations of the topologically connected Lie groups are equivalent
to the ordinary unitary representations of its unique maximal central
extension. This central extension is with respect to its first homotopy
group direct product with its second cohomology group. The second
cohomology group of the inhomogeneous symplect group is the one
dimensional abelian group and the central extension turns the $2n$
dimension abelian normal subgroup of translations of $\mathcal{I}\mathcal{S}p(
2n) $ into the Weyl-Heisenberg group. This is the quantum symmetry
origin of the Heisenberg commutation relations that are the Hermitian
representation of the Lie algebra of the Weyl-Heisenberg group.


\begin{thebibliography}{10}
\bibitem{wigner}Wigner, E. P. (1939). \textit{On the unitary representations of the inhomogeneous Lorentz group}. Annals of Math., \textbf{40}, 149--204 . 
\bibitem{Weinberg1}Weinberg, S. (1995).\textit{The Quantum Theory of Fields, Volume 1}. Cambridge: Cambridge.
\bibitem{bargmann}Bargmann, V. (1954). \textit{On Unitary Ray Representations of Continuous
Groups}. Annals Math., \textbf{59}, 1--46. 
\bibitem{mackey2}Mackey, G. W. (1958). \textit{Unitary Representations of Group Extensions. I}. Acta Math., \textbf{99}, 265--311. 
\bibitem{mackey}Mackey, G.W. (1976). \textit{The theory of unitary group representations}. Chicago: University of Chicago Press.
\bibitem{Azcarraga}Azcarraga, J.A., \& Izquierdo, J. M. (1998). \textit{Lie Groups, Lie Algebras, Cohomology and Some Applications in Physics}. Cambridge: Cambridge
University Press.
\bibitem{Low}Low, S.G. (2014). {\itshape Maximal quantum mechanical symmetry: Projective representations of the inhomogeneoussymplectic group}. J. Math. Phys. {\bfseries 55}, 022105.
\bibitem{barut}Barut, A. O., \& Raczka, R. (1986). \textit{Theory of Group Representations
and Applications}. Singapore: World Scientific.
\bibitem{Low13}Campoamor-Stursberg, R., \& Low, S. G. (2009). \textit{Virtual copies of semisimple Lie algebras in enveloping algebras of semidirect products and Casimir
operators}. J. Phys. A, \textbf{42}, 065205. 
\bibitem{folland}Folland, G. B. (1989). \textit{Harmonic Analysis on Phase Space}. Princeton:
Princeton University Press.

\end{thebibliography}
\end{document}